**Changes in lipid membranes may trigger amyloid toxicity in Alzheimer's disease.**

Elizabeth Drolle[1,2], Alexander Negoda[3], Keely Hammond[4], Evgeny Pavlov[3,5], and Zoya Leonenko[1,2,4]*

[1]Department of Biology, University of Waterloo, [2]Waterloo Institute of Nanotechnology, University of Waterloo, [3]Department of Physiology and Biophysics, Dalhousie University, Canada, [4]Department of Physics and Astronomy, University of Waterloo, [5]Department of Basic Sciences, New York University College of Dentistry, USA.

\* corresponding author: zleonenk@uwaterloo.ca

**Abstract**

**Amyloid-beta peptides (Aβ), implicated in Alzheimer's disease (AD), interact with the cellular membrane and induce amyloid toxicity. The composition of cellular membranes changes in aging and AD. We designed multi-component lipid models to mimic healthy and diseased states of the neuronal membrane. Using atomic force microscopy (AFM), Kelvin probe force microscopy (KPFM) and black lipid membrane (BLM) techniques, we demonstrated that these model membranes differ in their nanoscale structure and physical properties, and interact differently with Aβ. Based on our data, we propose a new hypothesis that changes in lipid membrane due to aging and AD may trigger amyloid toxicity through electrostatic mechanisms, similar to the accepted mechanism of antimicrobial peptide action. Understanding the role of the membrane changes as a key activating amyloid toxicity may aid in the development of a new avenue for the prevention and treatment of AD.**

**Introduction**

Alzheimer's disease (AD) is a progressive neurodegenerative disease which leads to severe impairment of memory and cognitive function and is characterized by the formation of amyloid-beta (Aβ) protein aggregates on neurons and cerebral blood vessels[1, 2]. While all amyloid aggregates, such as oligomers, fibrils, and plaques serve as cellular hallmarks of AD, small soluble oligomers have recently been shown to be more toxic to cells than larger fibrils[3]. There is currently no cure or prevention for AD; prospective strategies to prevent amyloid toxicity include inhibiting the formation of toxic oligomers, as well as preventing amyloid-damaging effect to the cellular membrane. In this work we propose and test a new hypothesis that changes in lipid membrane structure and properties may trigger amyloid toxicity.

It is known that Aβ aggregation occurs on the surfaces of neuronal cells, leading to amyloid plaque formation in the brain tissues of individuals diagnosed with AD[1, 2]. The cellular membrane is therefore recognized as a target for amyloid attack. Aβ-membrane interactions may occur through specific membrane receptors[4] as well as non-specifically with the lipid membranes themselves. Many studies have reported the effect of the membrane in general, and of lipid rafts on amyloid binding and toxicity[3, 5-13]. Despite these efforts, the molecular mechanism of amyloid toxicity remains unclear, which delays the development of a treatment for AD. Previous studies on the brain membrane lipid composition of AD patients have reveal changes in lipid composition that occur during disease progression. These include lowering the content of several types of phospholipids found in the inner leaflet of the membrane[14] and a decrease in sphingomyelin (SM) content due to increased sphingomyelinase activity[15]. Perhaps surprisingly, the role of these changes has not been investigated in relation to amyloid toxicity. One type of neuronal lipids – gangliosides - is of special interest, with some contradicting results as to what occurs to their levels as a result of AD. Reductions in the amount of gangliosides present in the membrane have been observed in several regions of AD brains compared to that of control brains[16-18] while other studies have suggested ganglioside plays a role in the formation of plaques and an increase in ganglioside monosialotetrahexosylganglioside (GM1) results in an increase of Aβ aggregation *in vitro*[19-21]. However, changes in membrane lipid composition may occur before the onset of AD symptoms and its corresponding cellular pathology. Recently, researchers demonstrated the predictive power of such changes in lipid composition in blood plasma as an early indicator of AD[22]. Changes in the composition of lipids found in blood plasma may be related to the changes in the lipid composition of neuronal membranes and /or membrane damage. Therefore it is of great interest to study the changes of structure and composition in neuronal membranes and their relevance to the amyloid-induced membrane damage, as these membrane changes may serve as an important switch to activate amyloid toxicity.

Biological cellular membranes are very complex and therefore model monolayers and bilayers are widely used to mimic the cellular membrane[23, 24]. While lipid models are very valuable for studying the mechanism of Aβ toxicity, earlier studies on model membranes cannot be easily related to *in vivo* animal and cellular studies, due to the fact that often, very simple models, composed of one or few lipid types, are used[5, 6, 8, 12, 13, 25-30]. Investigation of more complex model membranes will bridge our understanding of model systems and *in vivo* systems. In recent work, Sasahara et al. investigated the behaviour of Aβ in association with a lipid model containing five lipid constituents[31], and Bennett et al examined 29 neurolipidomic datasets and found evidence to support the idea of phospholipid metabolism having an important determinant in the conversion of AD[32]. Here, our goal is to not only increase membrane complexity to better mimic neuronal membrane, but most importantly to mimic healthy and AD states of neuronal cell membranes and elucidate the role of membrane changes in amyloid toxicity.

Currently, there are no lipid models mimicking healthy and AD neuronal membranes available in the literature, despite analysis of brain tissues showing changes in lipid composition with aging and AD[19, 33, 34]. Based on previous reports[31, 35, 36], we developed a membrane model that incorporates DPPC, POPC, sphingomyelin, cholesterol, and ganglioside GM1. These lipids are found in the outer leaflet of neuronal cell membranes[6, 19, 33, 37]. We hypothesize that changes in lipid composition of the AD brain affect the physical and structural properties of the neuronal cell membrane, compared to a healthy membrane, and that these changes in membrane fluidity, permeability, and lipid domain (raft) distribution, affect amyloid binding and make the neuronal membrane more susceptible to Aβ induced injury.

In order to test this hypothesis, we designed multicomponent lipid models that mimic healthy and diseased states of neuronal cell membrane, with the goal of elucidating structural differences between these models as well as the differences in Aβ binding and the damage that Aβ produces to the healthy versus AD model membranes. We used AFM and KPFM imaging to reveal the surface morphology and electrical surface potential of monolayers associated with these models. We used AFM imaging in liquid to visualize the binding of Aβ to the membrane as well as the BLM technique to monitor the membrane damage induced by Aβ upon binding.

## Materials and Methods

**Lipids.** 1,2-dipalmitoyl-*sn*-glycero-3-phosphocholine (DPPC), 1-palmitoyl-2-oleoyl-sn-glycero-3-phosphocholine (POPC), sphingomyelin (SM), cholesterol (Chol), and ganglioside monosialotetrahexosylganglioside (GM1) were purchased from Avanti Polar Lipids (Alabaster, AL) in powder form. Complex mixtures of these five constituents were made for analysis, and are outlined in Table 2. All other chemicals used were of reagent grade.

**Supported lipid monolayers.**

*Preparation via Langmuir-Blodgett monolayer technique.* Phospholipid monolayers were deposited on freshly cleaved mica (Ashville-Schoonmaker Mica Co., Newport News, VA) by the method of Langmuir-Blodgett (LB) deposition using a KSV-Nima LB microtrough (Biolin Scientific, Stockholm, Sweden). For sample preparation, solutions of lipid dissolved in chloroform at a concentration of 1 mg/mL (lipid/chloroform) were spread at the surface of the subphase and deposited on the mica substrates at a pressure of 35 mN/m with a dipper arm speed of 2 mm/min. The mica slide was allowed to air-dry for 10 minutes before being placed in a dessicator for a 24-hour period, prior to further analysis.

*AFM / KPFM Imaging.* AFM and KPFM imaging of monolayers supported on mica was performed using SmartSPM 1000 (AIST-NT) in air at room temperature and normal humidity using a MicroMasch gold coated cantilever (HQ:NSC14/Cr-Au) with a resonance frequency of 160 kHz and a spring constant of 5.0 N/m. KPFM imaging was performed in amplitude modulation mode (AM-KPFM) to achieve higher resolution and high sensitivity required for our biological samples than typical KPFM is capable of achieving. In this mode AM-KPFM imaging was done simultaneously with AFM imaging, AFM and AM-KPFM images of the sample correspond to the sample location[38].

*Data Processing and Analysis.* Data collected was processed using SPIP and AIST-NT image processing software. The AFM topography images were plane corrected by means of global leveling and global bow removal and filtered using noise reduction caused while scanning with high resolution Z-scale (picometers). KPFM images were not processed with any filters, to ensure the proper potential measurements; the raw data was used for average differences in electrical surface potential. Data was collected on 2 μm by 2 μm and 5 μm by 5 μm high-resolution images of monolayer samples. At least 10 images were analyzed for each sample. At least 3 samples were prepared for each membrane type. All quantitative results are presented as mean ± standard error of the mean (SEM), with significance determined using ANOVA tests. Any results determined to be significant are reported with a 95% confidence level.

**Aβ binding to the membrane.**

*Preparation of Lipid Bilayers.* Hydrated phospholipid bilayers were deposited on freshly cleaved mica (Ashville-Schoonmaker Mica Co., Newport News, VA) via vesicle fusion, as described in previous publications[12, 39]. Bilayers were covered with nanopure water and imaged in liquid using AFM.

*Aβ incubation on lipid membrane.* $Aβ_{1-42}$ (rPeptide, Bogart GA) was pretreated to ensure monomeric form according to Fezoui procedure[40]. Additional $Aβ_{1-42}$ was made in the lab of Dr. Paul Fraser (University of Toronto, Toronto ON) and was also studied to confirm and compare the findings obtained for the Aβ form rPeptide. Aβ was suspended in HEPES buffer (20 mM HEPES, 100 mM NaCl, pH 7.4) at a concentration of 40 mM (Aβ/buffer). 100 μL of the Aβ solution was added to pre-formed membranes and incubated for increasing time periods; at the end of the time period for each membrane, excess Aβ was gently rinsed away in order to stop the fibrilization process, with complete hydration maintained at all times. The membrane with Aβ deposits was kept in Nanopure water. At least two repeats of the incubations were completed for amyloid from each source, totaling to 4 trials for each time point.

*AFM imaging of lipid membrane in liquid.* AFM imaging of hydrated membrane and Aβ incubated membrane samples on a mica substrate was performed using Magnetic-Alternating-Current (MAC) mode on an Agilent AFM/SPM 5500 using Keysight Type II MAC mode rectangular cantilevers (force constant of 2.8 N/m and a resonance frequency in water of 30 kHz). Membrane imaging was conducted at ambient room temperature in liquid cell in Nanopure water, with hydration of the membrane maintained at all times throughout imaging.

*Data processing and analysis.* Data collected was processed using SPIP data processing software. The topography images were corrected via global levelling and global bow removal. Data was collected on 2 μm by 2 μm and 5 μm by 5 μm high-resolution images of membrane samples. All quantitative results are presented as mean ± SEM, with significant difference determined using ANOVA tests. Any differences determined to be significant are reported with a 95% confidence level. At least 3 samples were analyzed, each utilized a fresh batch of Aβ and lipid membranes, with at least 10 images taken for each sample from multiple locations on the membrane.

**Black Lipid Membrane (BLM) Studies.**

*Preparation of BLM samples.* Planar lipid bilayers were formed from a 15 mg/mL lipid solution in n-decane (Aldrich). The suspended bilayer was formed across the 200 μm aperture of a Delrin cup (Warner Instruments, Hamden, CT) by direct application of lipids[41], Both *cis* (voltage command side) and *trans* (virtual ground) compartments of the cup cuvette contained 150 mM NaCl, 2 mM $CaCl_2$, 10 mM Tris-HCl pH 7.4. 5 μM Aβ was added to the *cis* compartment of the cuvette. All measurements were performed at room temperature.

*Data recording and analysis.* Currents across lipid bilayers were recorded with a Planar Lipid Bilayer Workstation (Warner Instruments, Hamden, CT). The *cis* compartment was connected to the head stage input and the *trans* compartment was held at virtual ground via a pair of matched Ag/AgCl electrodes. Signals from voltage-clamped BLM were high-pass-filtered at 2.1 kHz using an eight-pole Bessel filter LPF-8 (Warner

Instruments), digitized (Data Translation digitizer) and recorded on PC using in-house analog-to-digital converter acquisition software developed by Elena Pavlova. For the statistical analysis data were averaged from at least three independent experiments and analyzed using Origin software. Experiments were performed in three separate trials for each sample. Each recorded trace was analyzed to get obtain the mean value of conductance. Results are expressed as mean ± SEM.

## Results

**Lipid Composition of Model Systems Mimicking Healthy and AD Membrane States.** Though there have been numerous studies on the interaction of lipid monolayers and membranes with Aβ[6, 8, 12, 13, 25-30, 42], many studies are carried out using simple models, consisting of one to three lipid types, which do not provide a good model for neuronal membrane. Based on previous studies on the composition of the outer leaflet of the neuronal membrane[6, 19, 33, 37], we designed three different multicomponent lipid models consisting of DPPC, POPC, sphingomyelin (SM), cholesterol (Chol), and ganglioside monosialotetrahexosylganglioside (GM1), which mimic healthy and AD neuronal membranes. The structure and properties of these lipids are shown in Table 1.

**Table 1: Structures and Properties of Lipids Studied.** This table outlines information about the five constituents of the models studied: DPPC, POPC, SM, Chol, and GM1. Phospholipid phase at ambient room temperature is indicated as samples were studied under these conditions. Dipole moment value is included due to its relevance in the KPFM study portion of this work. Lipid structures were adapted from Avanti Polar Lipids.

| Structure and Name | Abbreviation | Phase at 25°C | Phase Transition Temperature (°C) | Dipole Moment (D) |
|---|---|---|---|---|
| 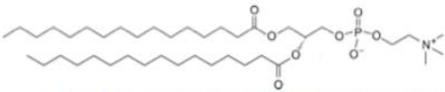 1,2-dipalmitoyl-sn-glycero-3-phosphocholine | DPPC | Gel | 41 | +0.82[43] |
| 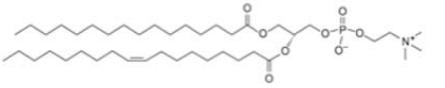 1-palmitoyl-2-oleoyl-sn-glycero-3-phosphocholine | POPC | Fluid | -2 | +0.473[44] |

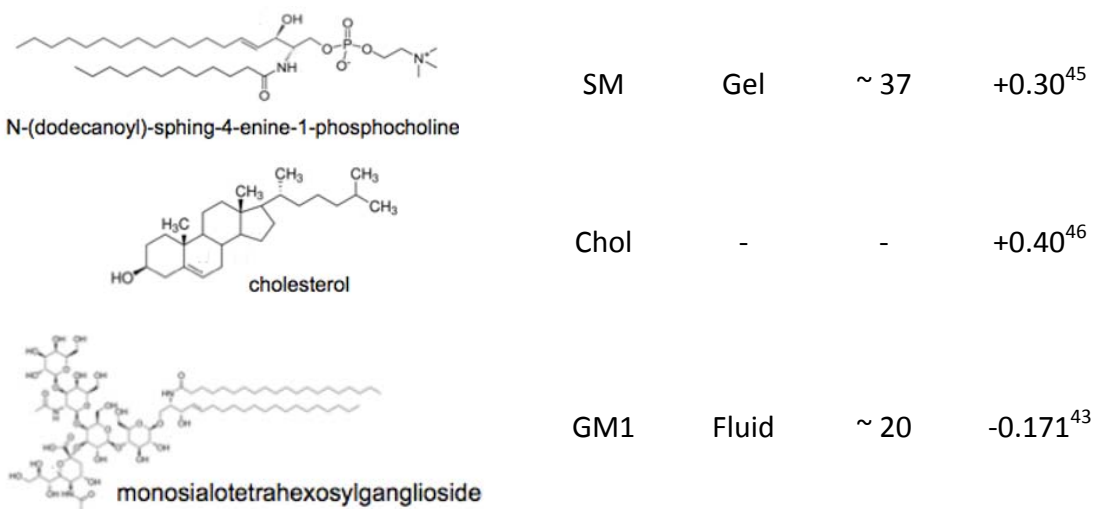

| | | | |
|---|---|---|---|
| SM | Gel | ~ 37 | +0.30[45] |
| Chol | - | - | +0.40[46] |
| GM1 | Fluid | ~ 20 | -0.171[43] |

Three models consist of varying lipid ratios of the same composition: DPPC-POPC-SM-Chol-GM1, shown in Table 2, and mimic healthy model membrane (HM), diseased model 1 accounting for decrease in GM1 content (D1) and diseased model 2 accounting for the decrease in both GM1 and SM content (D2). Such changes in membrane composition as a result of AD were observed *in vivo*[19, 33].

**Table 2: Complex lipid models mimicking healthy and AD neuronal membranes.** Lipid mixtures are all comprised of the same components but differ in their ratios (by weight) based on documented changes in membrane composition as a result of AD.

| Lipids | Ratio (by weight, %) | Ratio (by molarity, %) | Model Name |
|---|---|---|---|
| DPPC – POPC – SM – Chol – GM1 | 37 : 37 : 10 : 10 : 6 | 35.3 : 34.1 : 9.8 : 18.1 : 2.7 | ***Healthy Model*** |
| -mimics a "healthy" neuron | | | |
| DPPC – POPC – SM – Chol – GM1 | 39 : 39 : 10 : 10 : 2 | 36.5 : 35.2 : 9.7 : 17.8 : 0.09 | ***Diseased 1 (D1) Model*** |
| -mimics a neuron beginning to enter the "diseased" state with a decrease in the GM1 content (Diseased 1) | | | |
| DPPC – POPC – SM – Chol – GM1 | 42 : 42 : 4 : 10 : 2 | 39.4 : 38.0 : 3.9 : 17.8 : 0.09 | ***Diseased 2 (D2) Model*** |
| -mimics an increasingly "diseased" neuron with a decrease in both GM1 and SM content (Diseased 2) | | | |

The healthy model (Table 2) has five constituents commonly found in the outer leaflet of a general healthy neuronal cell membrane. Mass ratios were utilized to allow for easier comparison to our earlier studies[12, 39] and relative ratios were decided upon based on extrapolation from studies of lipid content in neuronal cells[47]. The D1 model was chosen to correlate with the reductions in gangliosides observed in membranes in several regions of AD brains compared to that of control brains, as well as with decreases in GM1 content as AD progresses[19, 29]. The D2 model has a decrease in both GM1 and SM compared to our model of a healthy neuronal membrane. This decreased amount of SM was chosen based on reported decrease in SM content due to increased sphingomyelinase activity in association with AD[15].

**Study of Monolayer Morphology and Electrical Surface Potential.** Using these combinations of lipids, we prepared supported monolayers of the three models and used AFM and KPFM in order to study the morphology and electrical surface potential distribution of each model monolayer. AFM allows for topographical imaging of a sample with nanoscale resolution, making it ideal for investigating the changes in monolayer morphology. KPFM is a variation of an AFM and has been shown useful for mapping electrical surface potential of a lipid monolayer at the nanoscale[38].

Figure 1 depicts the changes in topography (AFM), and electrical surface potential (KPFM), observed between the three models.

The healthy model (Figure 1A) shows the network of interconnected nanodomains spread across the monolayer. The topographical nanodomains have the average differences in height ($\Delta h$) of $0.986 \pm 0.024$ nm between higher and lower domains. The lateral dimensions of the higher domains ranged from 22.5 x 40.1nm up to over 200 x 52 nm. The lower domains were mostly small, less than 20 nm across, with few larger areas up to 150nm across. KPFM images of the monolayer corresponding to this model show some minor fluctuations in electrical surface potential (V) (Figure 1D), though no discernible patterns in $\Delta V$ are observed. The average roughness (variation in $\Delta V$ between higher and lower domains) of the sample is $24.^{6}4 \pm 1.10$ mV (as seen in Table 3).

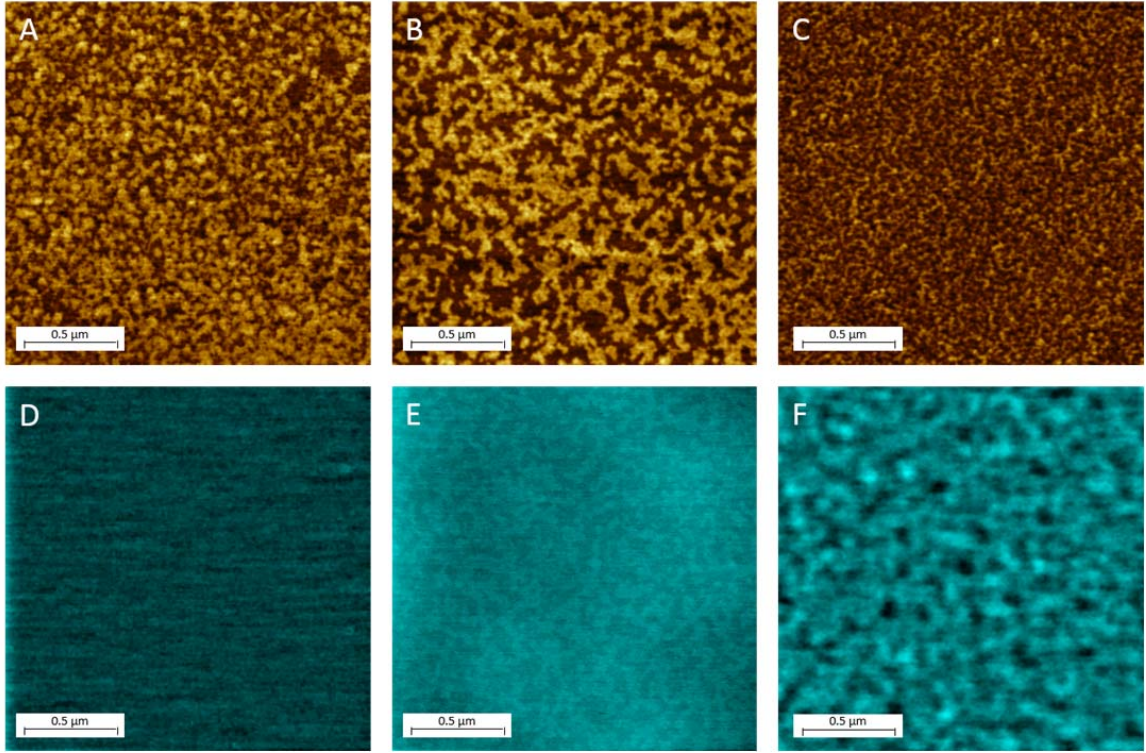

**Figure 1: Comparison of monolayer topography and electrical surface potential for three models.** AFM topography (in gold) and KPFM electrical surface potential (in blue) images are shown for healthy (A/D), diseased: D1 (B/E), and D2 (C/F) model systems in monolayer form. Cross sections of topography features are below each topography image. Scale bar denotes 500 nm.

In the AFM topography of the D1 model (Figure 1B), irregularly shaped interconnected higher and lower domains are observed. The higher domains are larger in area than in the healthy model, spanning up to 525 nm. The lower domains range from 35 to 200 nm across. The higher domains appear only slightly higher than those in the healthy model, with average Δh values of 1.051 ± 0.016 nm as compared to the healthy model (0.986 nm). The KPFM image of the D1 model (Figure 1E) shows organized, nanoscale electrostatic domains with the differences in electrical surface potential ΔV of 70.47 ± 5.41 mV, the highest average ΔV observed across the three samples. Topographic (AFM) and electrostatic (KPFM) nanodomains correlate with each other for the D1 model. Lower domains that are more disordered in nature are present in both the D1 and healthy model. However in the D1 model, these lower domains are larger in area than the healthy model; in the D1 model lower domains reached sizes of 200 nm across, whereas domains in the healthy model spanned up to 20 nm across. According to Connelly et al., Aβ is able to directly interact with a DOPC lipid bilayer and insert itself to form ion-conductive pores with an average outer diameter of 7.8 - 8.3 nm[48]. We expect that Aβ would most easily form pores in less ordered areas of the membrane corresponding with the regions in our monolayer models.

The D2 model (Figure 1C) shows a disruption of the larger, more ordered domains observed in the D1 model; it contains irregularly shaped features but smaller in area and more plentiful in number. They are quite narrow, with an average width of about 25 nm, and do not exceed 180 nm in length. The size of the lower domains are decreased when compared to the D1 model, with widths ranging from 27 to 40 nm compared with the widths of up to 200 nm seen in the D1 model. We also observe a drastic decrease in the average difference in height between higher and lower domains for the D2 model when compared to the other samples studied. The nanodomains observed in the D2 model had an average Δh of $0.500 \pm 0.03$ nm, which is significantly smaller than both the healthy and D1 models. This is likely due to the reduced SM content of this model, causing reduced lipid tail ordering. As seen from KPFM images (Fig 1F), the D2 model electrostatic domains present in the KPFM images did not correlate to AFM topography domains for this model (Figure 1C). An average ΔV of $11.63 \pm 0.59$ was measured for this model the smallest average ΔV observed between all the models. Results of the monolayer study are summarized in Table 3, which compares the average Δh and ΔV values for each of the three models studied.

**Table 3: Summary of Statistical Analysis of Mixed Lipid Monolayer Samples.** Average numbers for Δh (difference in height) and ΔV (difference in electrical surface potential), determined from AFM and KPFM images of monolayers corresponding to healthy, D1 and D2 models.

|  | Healthy Model | Diseased 1 Model | Diseased 2 Model |
|---|---|---|---|
| ***Topographical Domains*** | | | |
| Δh | $0.986 \pm 0.02$ nm | $1.051 \pm 0.016$ nm | $0.500 \pm 0.03$ nm |
| ***Features in Electrical Surface Potential*** | | | |
| ΔV | $24.64 \pm 1.10$ mV | $70.47 \pm 5.41$ mV | $11.63 \pm 0.59$ mV |

These results demonstrate that differences in topographical nanoheterogeneity and electrical surface potential are clearly seen in all three models (healthy, D1 and D2), which influence the interaction of each membrane with the charged Aβ peptide, discussed in the next section.

**Model Membranes and their Interactions with Aβ**.

**BLM Study: Amyloid Effect on Membrane Permeability.** We used Black Lipid Membrane (BLM) techniques to study the effects of Aβ on lipid bilayer conductance. This method allows for the measurement of ion currents across the membrane and membrane permeability to ions[49]. We compared currents measured across lipid bilayers of three different compositions corresponding to our model membranes shown in Table 2. These membranes were studied both in the absence and the presence of Aβ $_{(1-42)}$ peptide in order to investigate changes in membrane permeability caused by Aβ binding.

The conductance was observed in the control/"healthy" model without the addition of the Aβ (Figure 2A). The addition of 5 μM Aβ to the compartment of the cuvette containing the model led to an increase in noise amplitude with the following increase in conductance level. Although we recorded the increase in current, this increase was not statistically significant for the healthy model (Figure 2A Left panel, p=0.053, n=4). For the D1 membrane model, we found that the addition of Aβ caused a significant increase in the conductance of the membrane, which further increased with time. This effect is apparent from Figure 2B (p=5x10$^{-6}$, n=11), which illustrates an increase in conductance after 15 minutes of Aβ incubation compared to the conductance after 5 minutes of Aβ incubation. Finally, for the D2 model, we also observed that the addition of Aβ led to a significant increase (p=0.003, n=8) in the conductance of the membrane, which developed over the period of time. However, this increase was less than the increase in the conductance observed in the D1 model (Figure 2C). Overall, we established that the current across all of the tested membranes progressively increased with time reflecting membrane disintegration upon interaction with Aβ. Results of the quantitative comparison of the Aβ-induced conductance of tested model membranes are summarized in Figure 3.

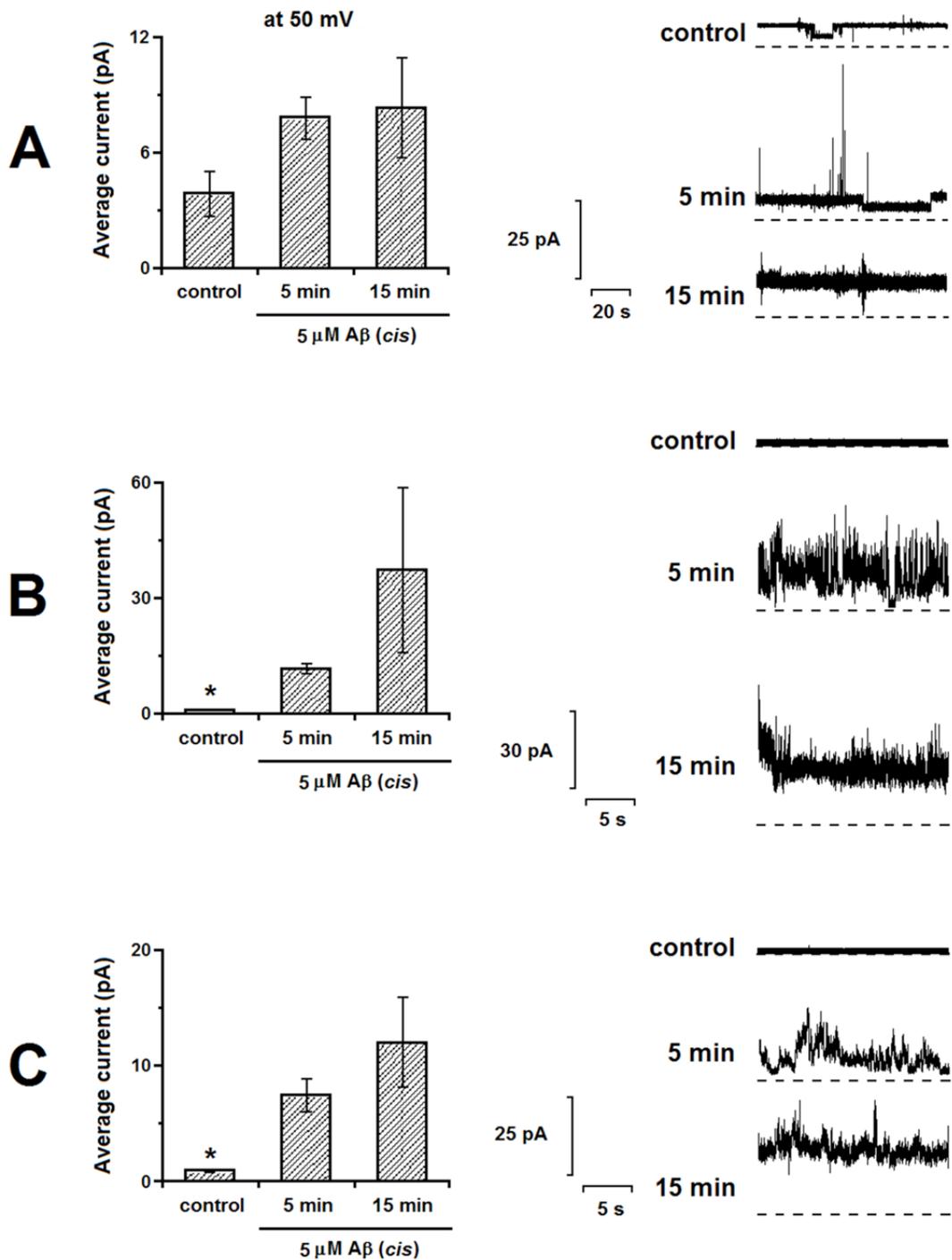

**Figure 2: Ion currents observed across the membrane: healthy model (A), D1 model (B), and D2 model (C).** Lipid membranes were suspended between symmetric aqueous solutions of 150 mM NaCl, 2 mM $CaCl_2$, 10 mM Tris pH 7.4. The left panel for each section shows average current at the voltage amplitude of 50 mV under control conditions (no additions), in 5 min after induction of conductance by the Aβ, in 15 min after induction of conductance by the Aβ. Representative currents at 50 mV are shown in the right panel. * indicates significantly lower current.

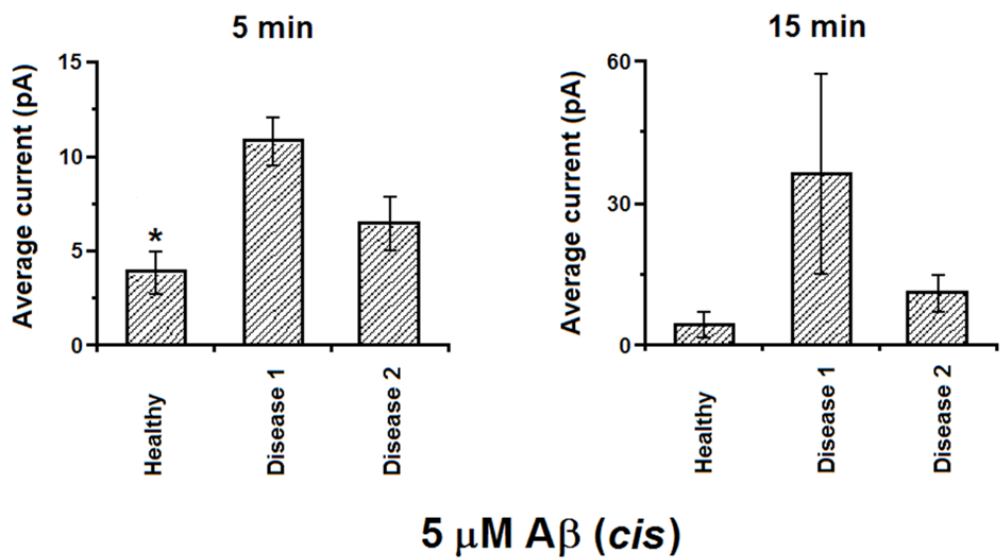

**Figure 3: Comparison of the currents induced by 5 μM of Aβ on different model membranes at voltage amplitude of 50 mV**. * indicates significantly lower current.

Of the three models, we found that in the presence of Aβ the D1 and D2 models both had higher conductance measurements than the healthy model. The highest current amplitude was observed in the D1 model membranes (in Fig 3, for 5 min (left): p=0.03, n=11; for 15 min (right): p=0.3, n=3).

Finally, we studied the interactions of Aβ with the membrane models using AFM, in order to determine how the nanoscale heterogeneity in topography and electrical surface potential observed in the monolayer study affects peptide-membrane interaction. Figure 4 illustrates schematically the presence of lipid domains (differing in height and electrical surface potential) in the complex multi-lipid membrane.

**AFM Study: Amyloid Incubation on Model Membrane Systems.** We formed supported membranes (bilayers) for each of the three lipid models and incubated solutions of Aβ in HEPES buffer in its monomeric form atop the membrane for 1, 4, 6, and 24 hours in buffer and imaged them with AFM in nanopure water, in order to maintain membrane hydration. We looked at four main factors: how Aβ binds to the membrane; the amount of Aβ binding and accumulating over time; the morphology of the Aβ aggregates on the membrane; and the presence of Aβ-induced membrane damage.

The differences in Aβ binding to the membrane and accumulation over time are shown in Figure 5 and results are summarized in Table 4.

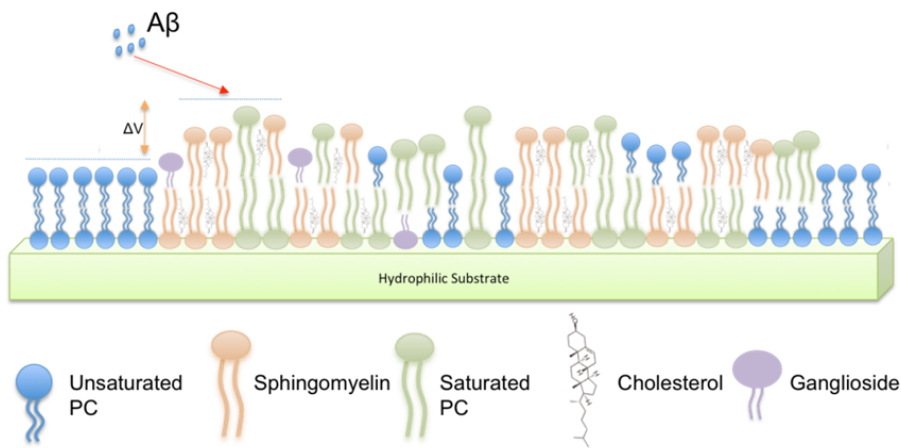

**Figure 4: Schematic of Aβ interacting with a model membrane (not to scale).** Arrangement of lipids present model bilayer system and phase separation leads to membrane nonhomogeneity, i.e., the presence of nanodomains, both topographical (Δh) and electrostatic (ΔV).

We followed the changes in the surface roughness with time which reflects Aβ binding to the membrane and accumulation of amyloid deposits on the membrane. In the healthy model, we observed the surface roughness increasing over time as the larger clusters are formed. This suggests that Aβ accumulation progressively increases with increasing incubation time. After 1 hour of incubation (Figure 5A) we see a uniform layer of aggregates randomly spread across the surface of the membrane, with a roughness of $0.267 \pm 0.05$ nm. The accumulation of Aβ aggregates on the membrane increase with time, with roughness increasing as incubation time increases: after 4 hours of incubation, average surface roughness was $0.536 \pm 0.11$ nm; after 6 hours of incubation (Figure 5D), roughness was $0.541 \pm 0.074$ nm; and after 24 hours of incubation, Aβ accumulation gave a surface roughness of $1.596 \pm 0.19$ nm. This is a progressive increase in the size of the Aβ clusters on the membrane.

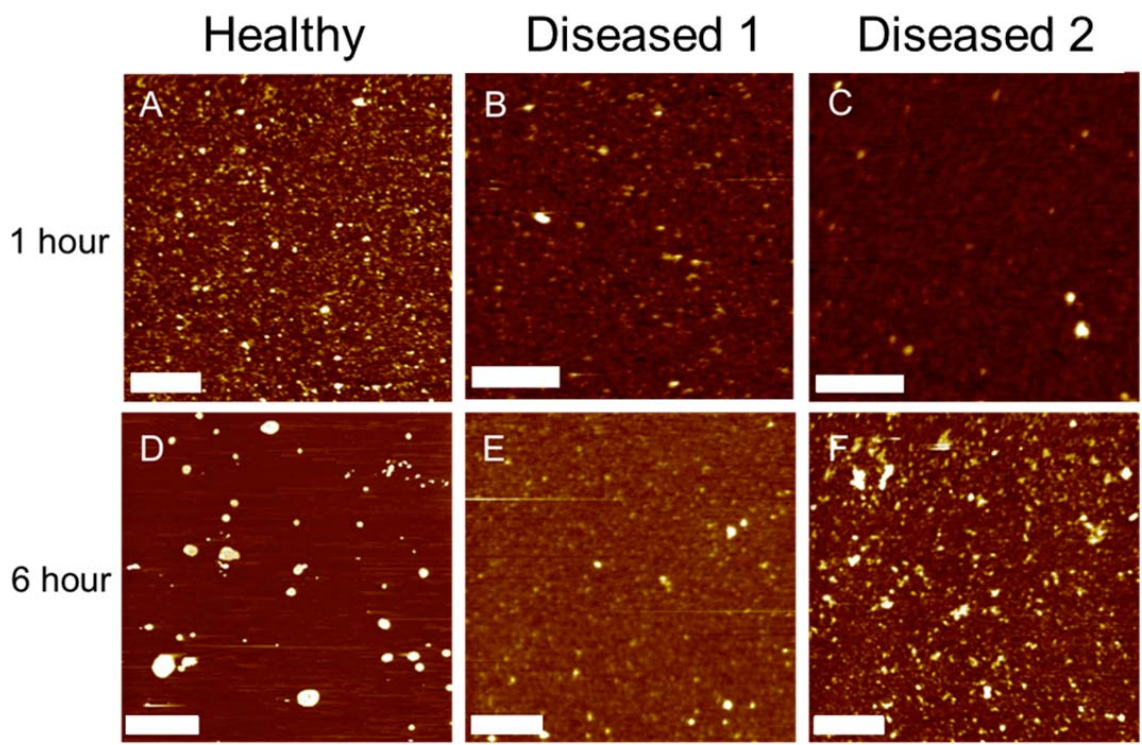

**Figure 5: Comparison of Aβ Incubation on Three Different Model Membranes for 1 and 6 hours.** AFM images in liquid illustrate the difference in Aβ accumulation between 1 and 6 hours of incubation time on a healthy model membrane (A and D respectively), a diseased 1 membrane (B and E) and a diseased 2 membrane (C and F). Image sizes are shown via the ruler scale bar in the left bottom hand corner of each image (the scale bar for A, D, E, and F corresponds to 1 μm; the scale bar for B and C corresponds to 500 nm).

**Table 4: Changes of Surface Roughness with Time due to Aβ Accumulation on the Membrane.** Aβ accumulation is quantified via surface roughness analysis for the HM, D1 and D2 models during 1h, 4h, 6h, and 24 h of incubation.

|  | Surface Roughness | | |
| --- | --- | --- | --- |
|  | Healthy Model (nm) | Diseased 1 Model (nm) | Diseased 2 Model (nm) |
| 1 Hour | 0.267 ± 0.05 | 0.487 ± 0.027 | 0.38 ± 0.033 |
| 4 Hour | 0.536 ± 0.11 | 0.42 ± 0.03 | 1.366 ± 0.12 |
| 6 Hour | 0.541 ± 0.074 | 0.57 ± 0.11 | 0.366 ± 0.045 |
| 24 Hour | 1.59 ± 0.19 | 0.44 ± 0.19 | 0.546 ± 0.026 |

In the D1 model, we saw a change in the accumulation pattern over time, in comparison to the healthy model. After 1 hour of incubation (Figure 5B), we observed a higher surface roughness in the D1 model, 0.487 ± 0.027 nm, than in the healthy model. This indicates more Aβ accumulation than in the healthy model. Although there is no discernible difference in the size and shape of Aβ species on the surface, in both models we saw small, spherical and irregularly shaped oligomers and aggregates. However, as time progressed, the roughness fluctuated: after 4 hours of incubation, we saw a decrease in surface roughness to 0.42 ± 0.03 nm; after 6 hours (Figure 5E), the surface roughness increased slightly to 0.57 ± 0.11 nm; and after 24 hours of incubation time, the average roughness decreased again to its lowest of all the time periods, to 0.44 ± 0.19 nm.

Finally, in the D2 model, we saw an initial large accumulation of Aβ species, indicated by an initial large increase in roughness, followed by a dramatic decrease. After 1 hour of incubation (Figure 5C), the surface roughness was 0.38 ± 0.03 nm, which almost quadrupled to 1.36 ± 0.12 nm after 4 hours. This indicates an initial large accumulation of Aβ on the surface of the membrane, without much membrane damage. However, after 6 hours (Figure 5F), we saw a large decrease in roughness of the model, with an average roughness of 0.366 ± 0.045 nm. At this point, Aβ clusters likely penetrated into the membrane, leading to a dramatic decrease in surface roughness, similar to what has been observed in fluid membranes[12, 50]. After 24 hours, the roughness increased slightly, to 0.546 ± 0.026 nm, suggesting a continuation of accumulation atop the Aβ-disrupted membrane.

**Discussion**

In this investigation, summarized in Table 5, we designed and studied three complex lipid models mimicking a healthy neuronal membrane, and two diseased states of the membrane (D1 and D2), mimicking changes in lipid compositions occurring in AD neurons. AFM and KPFM studies of monolayers show that the D1 and D2 models have different nanoscale surface morphologies (topographical domains and electrostatic domains) from each other and as well as compared with the healthy model.

**Table 5: Summary of Results.** AFM/KPFM study on topographical and electrical surface potential features of the models in monolayer form; Black Lipid Membrane analysis on the permeability of each model and the effect of Aβ on this conductance; and AFM liquid imaging of Aβ accumulation over time on each model membrane.

|  | Healthy Model | Diseased 1 Model | Diseased 2 Model |
|---|---|---|---|
| *Monolayer Morphology – AFM / KPFM Analysis* | | | |
| Topographical Δh | 0.986 ± 0.02 nm | 1.051 ± 0.016 nm | 0.500 ± 0.03 nm Lower than Healthy Model |
| Electrostatic Surface Potential ΔV | 24.64 ± 1.10 mV | 70.47 ± 5.41 mV Higher than Healthy Model | 11.63 ± 0.59 mV Lower than Healthy Model |
| *Black Lipid Membrane (BLM) Analysis* | | | |
| Pore Forming Activity (PFA) | No significant increase in PFA with addition of amyloid | Highest PFA of three systems studied Significant increase in PFA with addition of amyloid | Significant increase in PFA with addition of amyloid |
| *Aβ Binding to Model Membranes* | | | |
| Aβ Accumulation Over Time measured via Roughness Measurements | Increases with time | Fluctuate between increases and decreases over time Indicative of amyloid penetrating into membrane | Initial increase followed by large decrease Indicative of an initial accumulation event preceding membrane disruption/penetration |

Lipid domains originate from lipid separation commonly observed in multi-component lipid systems. Different lipids exist in different phases at ambient room temperature, such as liquid crystalline ($L_c$), liquid disordered ($L_d$), or with the presence of Chol, cholesterol-induced liquid-ordered phase domains, ($L_o$). These higher domains are likely to be rich in DPPC, SM, and Chol molecules, (Table 1), as well as GM1 molecules, known to associate with saturated phospholipids, SM, and Chol in lipid rafts. The lower domains are likely areas of high POPC concentration, as POPC is found in $L_d$ phase at room temperature. The changes in lipid composition in HM, D1 and D2 models, including GM1 and cholesterol, result in changes in membrane morphology, i.e. domain organization, size and ordering. These domains also differ in electrical surface potential.

We previously showed that similar nanoscale topographical and electrostatic domains are formed in a simple DOPC-Chol model and their presence causes preferential amyloid binding[39]. We showed that such changes in domain morphology and electrical surface potential in HM, D1 and D2 models influence their interaction with Aβ, and amyloid-induced damage.

Our BLM study shows that Aβ binds to the membrane and induces an increase in membrane conductance (ion permeability), which is a result of pore formation induced by amyloid. The significantly higher pore forming activity of the two diseased membrane models compared to the healthy model (Figure 2 and Table 5) supports the idea that the differences in membrane composition can have a strong effect on the interaction of the Aβ with the membrane and the extent to which the Aβ can cause damage and alteration of normal cell function by changing membrane permeability. This is consistent with different domain distribution and morphologies observed in each model with the monolayer studies.

Our AFM images illustrate Aβ binding to the membrane and indicate less penetration of Aβ into the healthy model membrane in good correlation with BLM results.

According to the proposed mechanisms of Aβ interaction with the membrane [50], depending on the membrane state, Aβ may either adsorb onto the surface of the lipid membrane (as seen in the healthy model) or partially penetrate into the membrane, causing membrane disruption and pore formation[48, 50]. The multiple types of membrane interaction that Aβ is capable of can be attributed to its complex charge distribution; this distribution allows for Aβ to bind to surfaces of varying charges and hydrophobicity[51].

Both the D1 and D2 models show higher penetration of Aβ into the membrane, inducing membrane damage, which correlates with the much higher membrane permeability recorded by BLM in D1 and D2 models. The differences in Aβ accumulation as well as differences in membrane disruption further show the significant effect of membrane lipid composition as well as nanoheterogeneity on its interaction with Aβ.

The importance of the effect of the composition of the membrane itself on amyloid - membrane interactions is of even greater interest as it has been shown recently that Aβ peptides share many similarities with antimicrobial peptides (AMP), which specifically recognize and kill bacterial cells through selective membrane-mediated recognition mechanism, causing disintegration of the bacterial membrane without affecting the host cell[52]. AMP and Aβ share common characteristics, including the ability to form fibrils in solution, capabilities of membrane interaction, ability to form ion channels and defects in the membrane[52]. *In vitro* studies have shown that Aβ has antimicrobial activity against eight common and clinically relevant microorganisms with a potency equivalent to or greater than AMP, which suggests the potential of Aβ is an unrecognized AMP of the innate immune system[53]. In fact, an exciting recent study showed the ability of Aβ to mediate the entrapment of unattached microbes in the brain, further suggesting that Aβ has protective roles in innate immunity [54]acting as an AMP in the brain.

AMPs are known to be very specific in recognizing the structure of bacterial membranes through electrostatic interactions. Similar to AMPs, Aβ is negatively charged (-3)[51] and thus may share this electrostatic mechanism. This allows Aβ to recognize changes in membrane structure and integrity through electrostatic interactions and the presence of electrostatic nanodomains in neuronal model membranes. This alteration in membrane composition and structure being a factor in the onset of AD also may help to explain why some people maintain normal cognitive abilities despite the presence of Aβ plaques[55].

**Conclusions**

In summary, we designed model lipid membranes which mimic the neuronal cell membrane in healthy and diseased states. We demonstrated that healthy and diseased model membranes differ in their nanoscale structure, which significantly influence the interaction of these membranes with Aβ (1-42). The diseased membrane models are more susceptible to interaction with Aβ and its damaging effects than a healthy membrane model. Based on our data and Aβ – AMP similarities reported in literature we propose a new hypothesis for the mechanism of amyloid toxicity in which the neuronal membrane changes play a crucial role: i.e. when neuronal cellular membrane changes in composition and properties due to aging or AD it becomes recognized by Aβ as foreign or "bacteria-like" membrane through electrostatic interactions and therefore undergo amyloid attack and disintegration. Therefore, sustaining neuronal cellular membrane in a healthy state may reduce the damaging effects of Aβ and serve as a preventative strategy against Alzheimer`s disease.

**Conflicts of interest**

The authors declare no competing financial interest.

**Acknowledgements**

The authors acknowledge contribution of Prof Paul Fraser and Ling Wu from University of Toronto, who provided Amyloid beta peptide for part of this study and critically read the manuscript. This work was supported by the Natural Science and Engineering Council of Canada (NSERC) grant [ZL], Heart and Stroke Foundation of Canada grant and American Heart Association grant [EP], an NSERC Canada Graduate Scholarship and WIN Fellowship [ED], and an NSERC Undergraduate Student Research Award [KH]. Technical support from AIST-NT and Keysight is greatly appreciated.